\documentclass[10pt,a4paper]{article}
\usepackage[a4paper,left=2.5cm,right=2.5cm,top=2.5cm,bottom=2.5cm]{geometry}
\usepackage[super,compress]{cite}
\usepackage{graphicx}
\usepackage{authblk}
\usepackage{xcolor}

\begin{document}

\title{General Expressions for $n_{\rm s}$ and $r$ for
Several Minimally Coupled Single Scalar Field Models}

\author[1]{Marko Stojanovic\thanks{marko.stojanovic@pmf.edu.rs}}
\author[2]{Dragoljub D. Dimitrijevic\thanks{ddrag@pmf.ni.ac.rs}}
\author[2]{Milan Milosevic\thanks{milan.milosevic@pmf.edu.rs}}

\affil[1]{Faculty of Medicine, University of Ni\v s, Serbia}
\affil[2]{Department of Physics, Faculty of Sciences and Mathematics, University of Ni\v s, Serbia}

\maketitle

\begin{abstract}
We derive a generalized expressions for the scalar power spectrum and the observational parameters of inflation (the scalar spectral index $n_{\rm s}$ and the tensor-to-scalar ratio $r$) for several single field scalar models minimally coupled with gravity. We work in the framework where the  perturbed Einstein equations, at linear order, are of the same form to those obtained in the standard \textit{k}-inflation, with generalized form of the pump field $z$. By generalizing the expressions for $z$ and the speed of sound squared $c_{\rm s}^2$, using the slow-roll approximation, we calculate $n_{\rm s}$ and $r$ up to the second order in the Hubble flow parameters. 
\end{abstract}


\section{Introduction}

Cosmological inflation [\citen{Linde:1981mu, Guth:1980zm}] have been considered in several different cosmological models with a scalar field. Various expressions for Lagrangian densities
are used: with canonical kinetic term (standard cosmology) [\citen{Liddle:2000cg}], with non-canonical kinetic term ($k$-cosmology [\citen{Armendariz-Picon:1999hyi}]) or just in non-standard form (tachyon cosmology and/or $k$-cosmology). We will discuss only minimally coupled models.

In addition to the standard cosmological models with tachyon scalar field, derived from the Friedmann equation that are obtained from the Einstein-Hilbert (EH) action [\citen{Steer:2003yu}], the model of tachyon inflation is also developed in cosmologies based on modified Friedmann (and Einstein) equations compared to their standard form: the cosmology obtained from the Randall-Sundrum II (RSII) model [\citen{Bilic:2017yll,Bilic:2016fgp,Dimitrijevic:2018nlc}] and the holographic cosmology [\citen{Bilic:2018uqx}]. Due to the modifications in Friedmann equation(s), the results for the observational parameters of inflation will differ from results for the standard inflationary models.
For the tachyon model in RSII cosmology [\citen{Bilic:2016fgp}] the authors calculated the scalar power spectrum 
taking into account the that the speed of sound squared is different from one, $c_{\rm s}^2\ne1$. For the tachyon model in holographic cosmology the scalar power spectrum have been calculated in an approximate scheme in which the modification of Einstein equations were treated as an approximate modification of the effective energy momentum tensor at the holographic boundary [\citen{Bilic:2018uqx}]. In this model, the tensor power spectrum is of the same form as in the standard model for tachyon inflation [\citen{Steer:2003yu}]. Besides, the complete perturbation theory in linear order for holographic model with a general \textit{k}-essence field have been developed in Ref.\ [\citen{Bertini:2020dvv}] by adjusting the formalism from Ref.\ [\citen{Garriga:1999vw}].

The differences in the expressions for the observational parameters, the scalar spectral index $n_{\rm s}$ and the tensor-to-scalar ratio $r$, in those models, come from the different expressions for the pump field $z$ and the speed of sound $c_{\rm s}$. Motivated by this results we calculate and obtain the generalized expressions for the scalar power spectrum and the parameters $n_{\rm s}$ and $r$ for the models with the single scalar field, minimally coupled with gravity. The perturbed Einstein equations at linear order are of the same form to those obtained in the standard \textit{k}-inflation, with the generalized form of the pump field $z$. 

The remainder of the paper is organized in five sections and two appendices. In Section 2 we briefly review main results from the perturbations in standard \textit{k}-inflation. In Section 3 by generalizing the expression for the pump field and the speed of sound, we calculate a generalized expressions for the scalar power spectrum and the parameters $n_{\rm s}$ and $r$. Then, we apply the obtained expressions, in Section 4, and reproduce the results from some models of inflation from the literature. Concluding remarks are given in Section 5, followed by Appendices A and B.

\section{Perturbations in \textit{k}-Inflation}\label{secstperturbation}

For the sake of completeness, in this section, we briefly review perturbations in \textit{k}-inflation, the class of inflationary models with
non-canonical kinetic term and non-standard expression for Lagrangian density. We will use the standard tachyon inflation as a typical example.

A general action for \textit{k}-inflation is of the form [\citen{Armendariz-Picon:1999hyi}]
\begin{equation}
S_{\rm matt}=\int d^4x\sqrt{-g}{\cal L}(X,\theta),
\end{equation}
where ${\cal L}={\cal L}(X,\theta)$ is an arbitrary function of the scalar field $\theta$ and its kinetic term $X\equiv g^{\mu\nu}\theta_{,\mu}\theta_{,\nu}$. The energy-momentum tensor associated with $S_{\rm matt}$ is
\begin{equation}
T_{\mu\nu}\equiv \frac{2}{\sqrt{-g}}\frac{\delta S_{\rm matt}}{\delta g^{\mu\nu}}=2{\cal L}_{,X}\theta_{,\mu}\theta_{,\nu}-g_{\mu\nu}{\cal L}.
\end{equation}
The subscript $,X$ denotes a partial derivative with respect to $X$. In the framework of fluid mechanics, it is straightforward to interpret physical quantities such as density, pressure and four-velocity in terms of scalar field quantities (see [\citen{Piattella:2013wpa}] for details). Using \textit{k}-essence [\citen{Armendariz-Picon:2000ulo}] fluid correspondence the energy-momentum tensor can be expressed in the form of
a perfect fluid
\begin{equation}
T_{\mu\nu}=(p+\rho)u_{\mu}u_{\nu}-pg_{\mu\nu},
\end{equation}
where $p$ and $\rho$ are pressure and energy density of the fluid defined through Lagrangian density $\mathcal{L}$ and its kinetic term $X$
\begin{equation}
p={\cal L},\quad \rho=2X{{\cal L}_{,X}-{\cal L}},
\label{pandrho}
\end{equation}
and $u_{\mu}\equiv\theta_{,\mu}/\sqrt{X}$ is four-velocity normalized as $u^{\mu}u_{\mu}=1$. Using the same correspondence the speed of sound can be defined
\begin{equation}
c_{\rm s}^2=\frac{p+\rho}{2X\rho_{,X}}.
\label{defcs}
\end{equation}

\subsection{Standard Tachyon Inflation}\label{}

For a fluid described by the tachyon field $\theta$ the Lagrangian density has the Dirac-Born-Infeld
(DBI) form [\citen{Sen:2002an}]
\begin{equation}
{\cal L}=-V(\theta)\sqrt{1-X},
\label{Ltachyon}
\end{equation}
where $V(\theta)$ is tachyonic potential. In the flat Universe, desribed by the Friedmann-Lemaitre-Robertson-Walker (FLRW) metric
\begin{equation}
ds^2=dt^2-a^2(t)d\vec{x}^2,
\end{equation}
where $a(t)$ is the scalar factor, the Friedmann equations are of the form
\begin{equation}
H^2=\frac{1}{3M_{\rm Pl}^{2}}\rho,
\label{Feq1}
\end{equation}
\begin{equation}
\dot{H}=-\frac{1}{2M_{\rm Pl}^{2}}(\rho+p),
\label{Feq2}
\end{equation}
where $M_{\rm Pl}=(8\pi G_{\rm N})^{1/2}$ and the Hubble parameter  is defined as $H(t)=\dot{a}/a$. The overdot denotes a derivative with respect to time. Combining (\ref{pandrho}), (\ref{Feq1}) and (\ref{Feq2}) leads to
\begin{equation}
\dot{\theta}^2=\frac{2}{3}\varepsilon_{1},
\label{dottheta}
\end{equation}
where $\varepsilon_{1}=-\dot{H}/H^2$ is the first Hubble flow parameter. From (\ref{defcs}) using (\ref{Ltachyon}) one finds $c_{\rm s}^2=1-X$. For a homogeneous tachyon field $\theta=\theta(t)$, 
 one finds, using (\ref{dottheta}), that the speed of sound 
\begin{equation}
c_{\rm s}^{2} = 1 - \frac{2}{3}\varepsilon_1,
\label{csstandard}
\end{equation}
is different from that of canonical single scalar field ($c_{\rm s}=1$).

\subsection{Scalar and Tensor Power Spectra}

In standard \textit{k}-inflation a scalar field $\theta$ is minimally coupled to gravity (EH action). The field equations are of the form
\begin{equation}
R_{\mu\nu}-\frac{1}{2}g_{\mu\nu}R=8\pi G_{\rm N}T_{\mu\nu},
\label{EHeq}
\end{equation}
where $g$ is the determinant of the FLRW metric and $R$ is the Ricci scalar, obtained from the Ricci curvature tensor $R_{\mu\nu}$. By introducing the small perturbations of the scalar field
\begin{equation}
\theta(t,\vec{x})=\theta(t)+\delta \theta(t,\vec{x}),
\end{equation}
and the perturbations of the spatially flat metric in the longitudinal gauge using  gravitational potential $\Phi$
\begin{equation}
ds^2=(1+2\Phi)dt^2-(1-2\Phi)a^2(t)(dr^2-r^2d\Omega^2),
\end{equation}
the perturbed Einstein equations at linear order, obrained from (\ref{EHeq}), can be conveniently expressed as the following set of equations
\begin{equation}
a\dot{\xi}=z^2c_{\rm s}^2\zeta,
\label{pEeq1}
\end{equation}
\begin{equation}
a\dot{\zeta}=z^{-2}\nabla^2\xi.
\label{pEeq2}
\end{equation}
The new variables $\zeta$ and $\xi$ are
\begin{equation}
\zeta=\Phi+H\frac{\delta\theta}{\dot{\theta}},\quad \xi=\frac{a\Phi}{4\pi G_{\rm N}},
\end{equation}
and the pump field $z$ is defined as
\begin{equation}
z = \frac{1}{\sqrt{4\pi G_{\rm N}}}\frac{a}{c_{\rm s}}\sqrt{\varepsilon_1}.
\label{defz}
\end{equation}
The quantity $\zeta$ measures the spatial curvature of comoving (or constant-$\theta$) hyper-surfaces.

The scalar power spectrum is $\mathcal{P}_{{\rm S}}(q)$ is defined for some gauge invariant scalar-type variable (fluctuations). The most suitable for inflationary period is to use scalar variable such as the comoving curvature perturbation $\zeta$, which is gauge invariant by construction. Then, the power spectrum reads
\begin{equation}
\mathcal{P}_{{\rm S}}(q) \equiv \frac{q^3}{2\pi^2}|\zeta_q|^2,
\label{P_HC1}
\end{equation}
\noindent where $\zeta_q$ is Fourier image of (gauge invariant scalar field) $\zeta $. Quantity $\zeta_q$ can be obtained with the help of (Fourier image of) the Mukhanov-Sasaki variable $v_q$
\begin{equation}
\zeta_q = \frac{1}{z}v_q.
\label{MS_variable}
\end{equation}
Note that $v_q$ is also gauge invariant scalar field. The time evolution of $v_q$ is given by Mukhanov-Sasaki equation
\begin{equation}
v^{\prime\prime}_q +\left(q^2c_{\rm s}^2-\frac{z^{\prime\prime}}{z}\right)v_q = 0,
\label{MS_equation}
\end{equation}
\noindent where prime denotes derivative with respect to conformal time $\eta$ defined in the Appendix A.

The tensor perturbations are related to the production of gravitational waves during inflation. In absence of anisotropic stress gravity waves are decoupled from matter. The tensor perturbation spectrum $\mathcal{P}_{{\rm T}}(q)$ in the first order of the Hubble flow parameters is given by the usual expression [\citen{Steer:2003yu, Bertini:2020dvv}]
\begin{equation}
\mathcal{P}_{{\rm T}}(q)\simeq \frac{16G_{\rm N}H^2}{\pi}[1-2(1+C)\varepsilon_{1}],
\label{PT}
\end{equation}
where $C=-2+\gamma +\ln 2\simeq -0.72$, and $\gamma$ is the Euler constant.

In this work, we assume that the tensor power spectrum is the same to expression (\ref{PT}) and there is no need to study the tensor perturbation in our set up. In the following section we are going to reconsider only scalar perturbation to obtain the scalar power spectrum.

\subsection{The Observational Parameters}

The scalar spectral index is defined as
\begin{equation}
n_{\rm s} - 1= \frac{d\ln \mathcal{P}_{{\rm S}}(q)|_{\rm HC}}{d\ln q},
\label{n_s}
\end{equation}
\noindent where $q$ is a (comoving) wave number of the primordial scalar-type fluctuations. Here, $\mathcal{P}_{{\rm S}}(q)$ is evaluated at the horizon crossing (HC), i.e.
\begin{equation}
\mathcal{P}_{{\rm S}}(q)|_{\rm HC} \equiv \mathcal{P}_{{\rm S}}(q)|_{qc_{\rm s}=aH },
\label{P_HC}
\end{equation}
\noindent where expression
\begin{equation}
qc_{\rm s}=aH,
\label{HC}
\end{equation}
\noindent defines the moment of horizon crossing. The ratio between the amplitude of tensor and scalar
perturbations is given by  the tensor-to-scalar ratio
\begin{equation}
r=\frac{\mathcal{P}_{{\rm T}}(q)|_{\rm HC}}{\mathcal{P}_{{\rm S}}(q)|_{\rm HC}}.
\label{defr}
\end{equation}
The parameters $n_{\rm s}$ and $r$, in terms of the Hubble flow (slow-roll) parameters $\varepsilon_{i}$ (introduced in the Appendix A),
in the standard cosmology are [\citen{Steer:2003yu}]
\begin{equation}
n_{\rm s}-1=-2\varepsilon_{1}-\varepsilon_{2}-\left[2\varepsilon_{1}^{2}+\left(2C+3-2\alpha\right)\varepsilon_{1}\varepsilon_{2}+C\varepsilon_{2}\varepsilon_{3}\right],
\label{nsst}
\end{equation}
\begin{equation}
r=16\varepsilon_{1}[1+C\varepsilon_{2}-2\alpha\varepsilon_{1}],
\label{rst}
\end{equation}
with $\alpha=0$ for the canonical scalar field inflation and $\alpha=1/6$ for tachyon inflation. At lowest order the predictions of the canonical scalar field and tachyon inflation are the same.

\section{Generalization}


The results from Section \ref{secstperturbation} are valid for the models  where the pump field is given by (\ref{defz}).
To extent consideration we start with a more general expressions for the pump field $z$ and for the speed of sound $c_{\rm s}$
\begin{equation}
z = \frac{1}{\sqrt{4\pi G_{\rm N}}}\frac{a}{c_{\rm s}}\sqrt{\varepsilon_1}\sqrt{A(H)},
\label{z_HBW}
\end{equation}
\begin{equation}
c_{\rm s}^{2} = 1 - B(H)\varepsilon_1.
\label{c_s_HBW}
\end{equation}
Here $A(H)$ and $B(H)$ are some functions of the Hubble parameter. This generalization is motivated by the results from the tachyon model
given in Ref.\ [\citen{Bilic:2018uqx}]
\begin{equation}
z = \frac{1}{\sqrt{4\pi G_{\rm N}}}\frac{a}{c_{\rm s}}\sqrt{\varepsilon_1}\sqrt{1-\frac{1}{2}\ell^2H^2},
\label{zhol}
\end{equation}
\begin{equation}
c_{\rm s}^{2} = 1 - \frac{4(2-\ell^2H^2)}{3(4-\ell^2H^2)}\varepsilon_1,
\label{cshol}
\end{equation}
as well as the results from Refs.\ [\citen{Bilic:2017yll}] and [\citen{Bertini:2020dvv}]. In (\ref{zhol}) and (\ref{cshol}) $\ell$ is an appropriate length scale.
In the limit $\ell H\rightarrow 0$ expressions (\ref{zhol}) and (\ref{cshol}) agree with (\ref{defz}) and (\ref{csstandard}), respectively. Our main motivation is to find $n_{\rm s}$ and $r$ as a function of $A(H)$ and $B(H)$ up to the second order in the Hubble flow parameters. In similar fashion, the parameters $n_{\rm s}$ and $r$ are calculated, to the first order in the Hubble flow parameters in Ref.\ [\citen{delCampo:2012qb}], for the model with modified Friedmann equation ${\cal F(H)}\sim \rho$, which will also be considered here.

\subsection{The Pump field and the Speed of Sound}

In order to find the expression for scalar power spectrum $\mathcal{P}_{{\rm S}}$ we need to solve equation (\ref{MS_equation}), for $z$ given by (\ref{z_HBW}). 
First, we need to find expression for $z^{\prime\prime}/z$ in terms of the Hubble flow parameters. Note that $\dot{A}(H) \sim \mathcal{O}(\varepsilon_{i})$, i.e.
\begin{equation}
\dot{A}(H) = -H^2\frac{dA(H)}{dH}\varepsilon_1.
\label{A_dot}
\end{equation}
We also need expression for derivative of $c_{\rm s}$ with respect to time $t$
\begin{equation}
\dot {c}_{\rm s} = - \frac{1}{2c_{\rm s}}M,
\label{c_s_dot}
\end{equation}
\noindent where
\begin{equation}
M \equiv -H^2\frac{dB(H)}{dH}\varepsilon_1^2 + HB(H)\varepsilon_1\varepsilon_2.
\label{M}
\end{equation}
\noindent Note that
\begin{equation}
M \sim \mathcal{O}(\varepsilon_{i}^2), \quad \dot {c}_{\rm s} \sim M + \mathcal{O}(\varepsilon_{i}^3),
\label{}
\end{equation}
\noindent which will be used, as the calculations are done up to the second order of the parameters $\varepsilon_i$.

We also define quantity
\begin{equation}
\mathcal{D} \equiv
\frac{c_{\rm s}}{a}\frac{d}{dt}\left(\frac{a}{c_{\rm s}}\right) +
\frac{1}{\sqrt{\varepsilon_1}}\frac{d\sqrt{\varepsilon_1}}{dt} +
\frac{1}{\sqrt{A(H)}}\frac{d\sqrt{A(H)}}{dt},
\label{D}
\end{equation}
\noindent which is connected with $z$
\begin{equation}
\frac{1}{z}\frac{dz}{dt} = \frac{\dot {z}}{z} = \mathcal{D}.
\label{D1}
\end{equation}
\noindent Taking derivative of (\ref{D1}) with respect to time $t$ we get
\begin{equation}
\frac{\ddot {z}}{z} = \mathcal{D}^2 + \dot {\mathcal{D}}.
\label{D2}
\end{equation}

Now, expression for $z^{\prime\prime}$ takes the form
\begin{equation}
z^{\prime\prime} = \frac{d}{d\eta}\left(\frac{dz}{d\eta}\right) =
\frac{d}{dt}\left(\frac{dz}{dt}\frac{dt}{d\eta}\right)\frac{dt}{d\eta},
\label{z_second}
\end{equation}
\noindent leading to
\begin{equation}
\frac{z^{\prime\prime}}{z} = \left(\mathcal{D}^2 + \dot {\mathcal{D}}\right)\left(\frac{dt}{d\eta}\right)^2 +
\mathcal{D}\frac{d}{dt}\left(\frac{dt}{d\eta}\right)\frac{dt}{d\eta}.
\label{z_second_over_z}
\end{equation}
\noindent Having in mind (\ref{dtau_deta4}) derived in the Appendix A we can rewrite expression (\ref{z_second_over_z}) as
\begin{equation}
\frac{z^{\prime\prime}}{z} =
\frac{a^2}{(1-\varepsilon_1^2-\varepsilon_1\varepsilon_2)^2}
\left(\mathcal{D}^2 + \dot {\mathcal{D}} + H\mathcal{D}\right).
\label{z_second_over_z1}
\end{equation}
Using (\ref{z_HBW}), (\ref{c_s_HBW}) and (\ref{A_dot}) we get (up to the second order in the Hubble flow parameters)
\begin{equation}
\mathcal{D} =
H\left\{1-\frac{1}{2}H\left(\frac{1}{A}\frac{dA}{dH}\right)\varepsilon_1 +
\frac{1}{2}\varepsilon_2 + \frac{1}{2H}M + \mathcal{O}(\varepsilon_{i}^3)\right\},
\label{D_final}
\end{equation}
\begin{eqnarray}
\mathcal{D}^2 &=&
H^2\left\{1 - H\left(\frac{1}{A}\frac{dA}{dH}\right)\varepsilon_1 + \varepsilon_2 +
\frac{1}{4}H^2\left(\frac{1}{A}\frac{dA}{dH}\right)^2\varepsilon_1^2 +
\frac{1}{4}\varepsilon_2^2\right\}
\nonumber\\
&-&
H^2\left\{\frac{1}{2}H\left(\frac{1}{A}\frac{dA}{dH}\right)\varepsilon_1\varepsilon_2\ -
\frac{1}{H}M + \mathcal{O}(\varepsilon_{i}^3) \right\},
\label{Dsquared_final}
\end{eqnarray}
\begin{eqnarray}
\dot {\mathcal{D}} &=&
H^2\left\{-\varepsilon_1+\left[H\left(\frac{1}{A}\frac{dA}{dH}\right) + \frac{1}{2}H^2\frac{1}{A}\frac{d^2A}{dH^2} -
\frac{1}{2}H^2\left(\frac{1}{A}\frac{dA}{dH}\right)^2\right]\varepsilon_1^2\right\}
\nonumber\\
&-&
H^2\left\{\left[\frac{1}{2} + \frac{1}{2}H\left(\frac{1}{A}\frac{dA}{dH}\right)\right]\varepsilon_1\varepsilon_2 -
\frac{1}{2}\varepsilon_2\varepsilon_3 + \mathcal{O}(\varepsilon_{i}^3)\right\},
\label{D_dot_final}
\end{eqnarray}
\noindent where we also used (several times) expressions (\ref{h_dot})-(\ref{e2_dot}) derived in the Appendix A as well as (\ref{c_s_dot}).
This leads to the (prefinal) expression
\begin{eqnarray}
\frac{z^{\prime\prime}}{z} &=& \frac{a^2H^2}{(1-\varepsilon_1^2-\varepsilon_1\varepsilon_2)^2}
\nonumber\\
&\times &
\left\{
2 - \left[1+\frac{3}{2}H\left(\frac{1}{A}\frac{dA}{dH}\right)\right]\varepsilon_1 + \frac{3}{2}\varepsilon_2 +
A_1\varepsilon_1^2 + A_2\varepsilon_1\varepsilon_2 + \frac{1}{4}\varepsilon_2^2 + \frac{1}{2}\varepsilon_2\varepsilon_3
\right\},
\label{z_second_over_z_prefinal}
\end{eqnarray}
\noindent where
\begin{equation}
A_1(H) \equiv H\left(\frac{1}{A}\frac{dA}{dH}\right) + \frac{1}{2}H^2\frac{1}{A}\frac{d^2A}{dH^2} -
\frac{1}{4}H^2\left(\frac{1}{A}\frac{dA}{dH}\right)^2-\frac{3}{2}H\frac{dB}{dH},
\label{B1}
\end{equation}
\begin{equation}
A_2(H) \equiv \frac{1}{2} - H\left(\frac{1}{A}\frac{dA}{dH}\right) +\frac{3}{2}B-1.
\label{B2}
\end{equation}
If we use series expansion up to the second order
\begin{equation}
\frac{1}{(1-\varepsilon_1^2-\varepsilon_1\varepsilon_2)^2} = 1 + 2\varepsilon_1^2 + \varepsilon_1\varepsilon_2 + \mathcal{O}(\varepsilon_{i}^3),
\label{expansion}
\end{equation}
\noindent then expression (\ref{z_second_over_z_prefinal}) receives the form
\begin{eqnarray}
\frac{z^{\prime\prime}}{z} &=& a^2H^2
\left\{
2 - \left[1+\frac{3}{2}H\left(\frac{1}{A}\frac{dA}{dH}\right)\right]\varepsilon_1 + \frac{3}{2}\varepsilon_2\right\}
\nonumber\\
&+&
a^2H^2
\left\{
\left[4+A_1\right]\varepsilon_1^2 + \left[2+A_2\right]\varepsilon_1\varepsilon_2 + \frac{1}{4}\varepsilon_2^2 +
\frac{1}{2}\varepsilon_2\varepsilon_3 + \mathcal{O}(\varepsilon_{i}^3)
\right\}.
\label{z_second_over_z_prefinal1}
\end{eqnarray}
To solve equation (\ref{MS_equation}) we need to rewrite (\ref{z_second_over_z_prefinal1}) in explicit (conformal) time dependence form. Details are contained in the Appendix B.

\subsection{Asymptotic Solution of Mukhanov-Sasaki equation}

Mukhanov-Sasaki equation (\ref{MS_equation}) is Bessel-type differential equation if
\begin{equation}
\frac{z^{\prime\prime}}{z} = \frac{1}{\eta^2}\left(\nu^2 - \frac{1}{4}\right),
\label{z_second_over_z_prefinal2}
\end{equation}
\noindent holds, i.e. if it is rewritten in the form
\begin{equation}
v^{\prime\prime}_q +\left[q^2c_{\rm s}^2-\frac{1}{\eta^2}\left(\nu^2 - \frac{1}{4}\right)\right]v_q = 0.
\label{MS_equation1}
\end{equation}
\noindent Here, $\nu$ is real quantity that need to be determined. We do not need to solve it explicitly, we just need to know particular
asymptotic expression of the solution (squared), which is of the form
\begin{equation}
v_q^2(\eta) \simeq \left[2^{\nu-3/2}\frac{\Gamma(\nu )}{\Gamma(3/2)}\right]^2
\frac{1}{2qc_{\rm s}}\left(-qc_{\rm s}\eta\right)^{2(-\nu+1/2)}.
\label{v_squared}
\end{equation}
\noindent This (particular) asymptotic solution is valid for slowly varying $\nu$ and $c_{\rm s}$ as a function of conformal time $\eta$,
and this solution corresponds to the fluctuations on the superhorizon scale
\begin{equation}
qc_{\rm s} \ll aH.
\label{superhorizon}
\end{equation}
It is important to note that $\nu$ is not a small quantity, while $(\nu - 3/2)$ is
\begin{equation}
(\nu - \frac{3}{2}) \sim \mathcal{O}(\varepsilon_{i}).
\label{1}
\end{equation}
%
Having in mind this we can write
\begin{equation}
\left(-qc_{\rm s}\eta\right)^{2(-\nu+1/2)} =
\frac{1}{(-qc_{\rm s}\eta)^2}\left(-qc_{\rm s}\eta\right)^{-2(\nu-3/2)} \simeq
\frac{1}{(-qc_{\rm s}\eta)^2}.
\label{2}
\end{equation}
\noindent Using (\ref{2}) and (\ref{eta_SR1}) derived in the Appendix A, solution (\ref{v_squared}) becomes
\begin{equation}
v_q^2(\eta) \simeq F \frac{1}{2qc_{\rm s}}\left(\frac{aH}{2qc_{\rm s}}\right)^2
\frac{1}{(1+\varepsilon_1)^2},
\label{v_squared1}
\end{equation}
\noindent where we introduce
\begin{equation}
F \equiv \left[2^{\nu-3/2}\frac{\Gamma(\nu )}{\Gamma(3/2)}\right]^2.
\label{F}
\end{equation}
Having in mind (\ref{1}), we can use approximate expressions
\begin{equation}
2^{\nu-3/2} \simeq 1 + (\nu-\frac{3}{2})\ln 2,
\label{3}
\end{equation}
\begin{equation}
\frac{\Gamma(\nu )}{\Gamma(3/2)} \simeq 1 - (\nu-\frac{3}{2})(C + \ln 2).
\label{4}
\end{equation}
Object $F$ now receives the form
\begin{equation}
F = 1-2C(\nu-\frac{3}{2}) + (C^2-2\tilde{C})(\nu-\frac{3}{2})^2 + \mathcal{O}(\varepsilon_{i}^3),
\label{F1}
\end{equation}
\noindent where $\tilde{C}=(C+\ln 2)\ln 2$, and the solution becomes
\begin{equation}
v_q^2(\eta) \simeq \frac{1}{2qc_{\rm s}}\left(\frac{aH}{qc_{\rm s}}\right)^2
\frac{1}{(1+\varepsilon_1)^2}
\left\{1-2C(\nu-\frac{3}{2}) + (C^2-2\tilde{C})(\nu-\frac{3}{2})^2\right\}.
\label{v_squared2}
\end{equation}
We are now ready to proceed with the scalar power spectrum (\ref{P_HC1}).

\subsection{The Scalar Power Spectrum}

Using (\ref{MS_variable}), (\ref{z_HBW}) and (\ref{v_squared2}) the expression (\ref{P_HC1}) for the scalar power spectrum at the moment of horizon crossing becomes
\begin{equation}
\mathcal{P}_{{\rm S}}(q)|_{\rm HC} = \frac{G_{\rm N}}{\pi}
\frac{H^2}{c_{\rm s}A(H)}
\frac{1}{\varepsilon_1(1+\varepsilon_1)^2}
\left\{1-2C(\nu-\frac{3}{2}) + (C^2-2\tilde{C})(\nu-\frac{3}{2})^2\right\}|_{\rm HC}.
\label{P_HC2}
\end{equation}
Taking logarithm and using the horizon crossing condition (\ref{HC}) to define
\begin{equation}
d\ln q = H\left(1 - \varepsilon_1 + \frac{1}{H}M + \mathcal{O}(\varepsilon_{i}^3)\right)dt
\simeq H(1 - \varepsilon_1)dt,
\label{d_lnk}
\end{equation}
\noindent i.e.
\begin{equation}
\frac{d}{d\ln q} = \frac{1}{H(1 - \varepsilon_1)}\frac{d}{dt},
\label{d_lnk1}
\end{equation}
\noindent we get
\begin{equation}
\frac{d\mathcal{P}_{{\rm S}}(q)|_{\rm HC}}{d\ln q} =
\frac{1}{H(1 - \varepsilon_1)}
\left\{
\frac{2}{h}\dot{H} - \frac{1}{c_{\rm s}}\dot{c}_{\rm s} - \frac{1}{A}\dot{A} -
\frac{1}{\varepsilon_1}\dot{\varepsilon}_1 - \frac{2}{1+\varepsilon_1}\dot{\varepsilon}_1 + \frac{1}{F}\dot{F}
\right\}.
\label{P_HC3}
\end{equation}
\noindent With the help of (\ref{c_s_dot}) and (\ref{h_dot})-(\ref{e2_dot}) derived in the Appendix A expression (\ref{P_HC3}) is rewritten
\begin{equation}
\frac{d\mathcal{P}_{{\rm S}}(q)|_{\rm HC}}{d\ln q} =
\frac{1}{1 - \varepsilon_1}
\left\{
-\left[2-H\left(\frac{1}{A}\frac{dA}{dH}\right)\right]\varepsilon_1 - \varepsilon_2 -
2\frac{\varepsilon_1\varepsilon_2}{1+\varepsilon_1} + \frac{1}{2H}M + \frac{1}{HF}\dot{F}
\right\}.
\label{P_HC4}
\end{equation}
\noindent The third and the last terms in (\ref{P_HC4}) can be written as
\begin{equation}
2\frac{\varepsilon_1\varepsilon_2}{1+\varepsilon_1} = 2\varepsilon_1\varepsilon_2 + \mathcal{O}(\varepsilon_{i}^3),
\label{5}
\end{equation}
\begin{equation}
\frac{1}{HF}\dot{F} = -\frac{2C}{H}\frac{d}{dt}(\nu - \frac{3}{2}) + \mathcal{O}(\varepsilon_{i}^3),
\label{6}
\end{equation}
\noindent where we used (\ref{F1}). From (\ref{1}) we also conclude that
\begin{equation}
\frac{d}{dt}(\nu - \frac{3}{2}) \sim \mathcal{O}(\varepsilon_{i}^2).
\label{7}
\end{equation}
%
Using expansion
\begin{equation}
\frac{1}{1-\varepsilon_1} = 1 + \varepsilon_1 + \varepsilon_1^2 + \mathcal{O}(\varepsilon_{i}^3),
\label{8}
\end{equation}
\noindent expression (\ref{P_HC4}) can be written more explicitly
\begin{eqnarray}
\frac{d\mathcal{P}_{{\rm S}}(q)|_{\rm HC}}{d\ln q} =
&-& \left[2-H\left(\frac{1}{A}\frac{dA}{dH}\right)\right]\varepsilon_1 - \varepsilon_2 -
\left[2-H\left(\frac{1}{A}\frac{dA}{dH}\right) + \frac{1}{2}H\frac{dB}{dH}\right]\varepsilon_1^2
\nonumber\\
&-& \left[3-\frac{1}{2}B\right]\varepsilon_1\varepsilon_2 -
\frac{2C}{H}\frac{d}{dt}(\nu - \frac{3}{2}).
\label{P_HC5}
\end{eqnarray}

\subsection{Generalized Expressions for $n_{\rm s}$ and $r$}

Expression (\ref{n_s}) for the scalar spectral index, using (\ref{P_HC5}) now reads
\begin{eqnarray}
n_{\rm s} - 1 =
&-& \left[2-H\left(\frac{1}{A}\frac{dA}{dH}\right)\right]\varepsilon_1 - \varepsilon_2 -
\left[2-H\left(\frac{1}{A}\frac{dA}{dH}\right) + \frac{1}{2}H\frac{dB}{dH}\right]\varepsilon_1^2
\nonumber\\
&-& \left[3-\frac{1}{2}B\right]\varepsilon_1\varepsilon_2 -
\frac{2C}{H}\frac{d}{dt}(\nu - \frac{3}{2}).
\label{n_s1}
\end{eqnarray}
The last term can be transformed as follow. First, we can write
\begin{equation}
(\nu - \frac{3}{2}) = b_1\varepsilon_1 + b_2\varepsilon_2 + \dots\;,
\label{9}
\end{equation}
\noindent up to the second order in the Hubble flow parameters, where $b_i$ are still unknown
coefficients and are functions of $H$. Using (\ref{h_dot}) derived in the Appendix A we can write
\begin{equation}
\frac{d}{dt}b_i(H) = -H^2\frac{db_i}{dH}\varepsilon_1.
\label{b_dot}
\end{equation}
The next expression also holds
\begin{equation}
\frac{d}{dt}(\nu - \frac{3}{2}) =
H\left\{
-H\frac{db_1}{dH}\varepsilon_1^2 + \left[b_1-H\frac{db_2}{dH}\right]\varepsilon_1\varepsilon_2 +
b_2\varepsilon_2\varepsilon_3 + \mathcal{O}(\varepsilon_{i}^3)
\right\},
\label{10}
\end{equation}
\noindent and can rewrite (\ref{n_s1}) as
\begin{eqnarray}
n_{\rm s} - 1 =
&-& 2\left[1-\frac{1}{2}H\left(\frac{1}{A}\frac{dA}{dH}\right)\right]\varepsilon_1 - \varepsilon_2
\nonumber\\
&-& 2\left[1-\frac{1}{2}H\left(\frac{1}{A}\frac{dA}{dH}\right) - CH\frac{db_1}{dh} + \frac{1}{4}H\frac{dB}{dH}\right]\varepsilon_1^2
\nonumber\\
&-& \left[3-\frac{1}{2}B + 2Cb_1 - 2CH\frac{db_2}{dH}\right]\varepsilon_1\varepsilon_2 -
2Cb_2\varepsilon_2\varepsilon_3.
\label{n_s2}
\end{eqnarray}
For the sake of completeness, the calculation of the coefficients $b_i$ is presented in the Appendix B.

Using $b_1$ and $b_2$ from (\ref{nu_squared2}) derived in the Appendix B we can rewrite expression (\ref{n_s2}) for scalar spectral index
\begin{eqnarray}
n_{\rm s} - 1 =
&-& 2\left[1-\frac{1}{2}H\left(\frac{1}{A}\frac{dA}{dH}\right)\right]\varepsilon_1 - \varepsilon_2
\nonumber\\
&-& 2\left[1-\frac{1}{2}H\left(\frac{1}{A}\frac{dA}{dH}\right) - CH\frac{db_1}{dH} + \frac{1}{4}H\frac{dB}{dH}\right]\varepsilon_1^2
\nonumber\\
&-& \left[3-\frac{1}{2}B + 2C-2C\frac{1}{2}H\left(\frac{1}{A}\frac{dA}{dH}\right)\right]\varepsilon_1\varepsilon_2 - C\varepsilon_2\varepsilon_3,
\label{n_s3}
\end{eqnarray}
where
\begin{equation}
\frac{db_1}{dH}=\frac{d}{dH}\left[1-\frac{1}{2}H\left(\frac{1}{A}\frac{dA}{dH}\right)\right].
\label{b1_dot}
\end{equation}
This is the final expression for the scalar spectral index up to the second order with respect to the Hubble flow parameters.

Using the definition for the tensor-to-scalar ratio, equation (\ref{defr}), together with (\ref{PT}) and (\ref{P_HC2}) we obtain
\begin{equation}
r= 16A(H)c_{\rm s}\varepsilon_{1}(1+\varepsilon_{1})^{2}\frac{1-2(1+C)\varepsilon_{1}}{1-2C(\nu-\frac{3}{2}) + (C^2-2\tilde{C})(\nu-\frac{3}{2})^2},
\end{equation}
i.e.
\begin{equation}
r\simeq 16A(H)c_{\rm s}\varepsilon_{1}(1+\varepsilon_{1})^{2}\left(1-2(1+C)\varepsilon_{1}\right)\left(1+2C(\nu-\frac{3}{2})+ \mathcal{O}(\varepsilon_{i}^2)\right).
\end{equation}
The expressions for $r$ can be put in the form
\begin{equation}
r= 16A\varepsilon_{1}\left[1-\left(CH\left(\frac{1}{A}\frac{dA}{dH}\right)+\frac{B}{2}\right)\varepsilon_{1}+C\varepsilon_{2}\right],
\label{finalr}
\end{equation}
where we used (\ref{nu_final})
derived in the Appendix B and approximate expression for the speed of sound  $c_{\rm s}\simeq 1-1/2B\varepsilon_{1}$.
Expressions (\ref{n_s3}) and (\ref{finalr}) for $n_{\rm s}$ and $r$ up to the second order with respect to the Hubble flow parameters are our main results.

As we shall shortly demonstrate in the next section, by a straightforward calculation, that the generalized expression for the observational parameters, expressions (\ref{n_s3}) and (\ref{finalr}), for given $z$ and $c_{\rm s}^2$ lead to the same results for several models studied in Refs.\ [\citen{Steer:2003yu,Bilic:2016fgp,Bilic:2017yll,Bilic:2018uqx,Bertini:2020dvv,Dimitrijevic:2018nlc}] and [\citen{delCampo:2012qb}].

\section{Checking Obtained Expressions}

In this section we apply the generalized expressions for the scalar spectral index and the tensor-to-scalar ratio, (\ref{n_s3}) and (\ref{finalr}), to several inflationary model, and confront it with existing results.

\subsection{Inflation with Tachyon Scalar Field}

In the standard tachyon inflation  [\citen{Steer:2003yu}] the perturbed Einstein equations can be rewrite as (\ref{pEeq1}) and (\ref{pEeq2}), where the pump field is of the form
\begin{equation}
z = \frac{1}{\sqrt{4\pi G}}\frac{a\sqrt{\varepsilon_1}}{c_{\rm s}},
\label{ztachyon}
\end{equation}
and $c_{\rm s}^2$ is given by (\ref{csstandard}). The form of $z$ and $c_{\rm s}^2$ imply to take $A=1$ and $B=2/3$ in (\ref{n_s3}) and (\ref{finalr}), yielding
\begin{equation}
n_{\rm s}-1=-2\varepsilon_{1}-\varepsilon_{2}-\left[2\varepsilon_{1}^{2}+\left(2C+\frac{8}{3}\right)\varepsilon_{1}\varepsilon_{2}+C\varepsilon_{2}\varepsilon_{3}\right],
\end{equation}
\begin{equation}
r=16\varepsilon_{1}[1+C\varepsilon_{2}-\frac{1}{3}\varepsilon_{1}],
\end{equation}
that agrees with (\ref{nsst}) and (\ref{rst}) for $\alpha=1/6$.

\subsection{Inflation with Canonical Scalar Field}

In canonical scalar field inflation in standard cosmology [\citen{Armendariz-Picon:1999hyi}] field the $z$ is of the same form as (\ref{ztachyon}) with $c_{\rm s}=1$, i.e. $A=1$ and $B=0$. We obtain
\begin{equation}
n_{\rm s}-1=-2\varepsilon_{1}-\varepsilon_{2}-\left[2\varepsilon_{1}^{2}+\left(2C+3\right)\varepsilon_{1}\varepsilon_{2}+C\varepsilon_{2}\varepsilon_{3}\right],
\label{nsssfi}
\end{equation}
\begin{equation}
r=16\varepsilon_{1}[1+C\varepsilon_{2}],
\label{rssfi}
\end{equation}
that agrees with (\ref{nsst}) and (\ref{rst}) for $\alpha=0$. Obviously, the different value of $B$ in canonical and tachyon inflation in standard cosmology is the reason why $n_{\rm s}$ and $r$ are the same up to the first order in the Hubble flow parameters. In addition, using expression (\ref{z_second_over_z_prefinal1}), we can verify that the value of expression $z^{\prime\prime}/z$ in standard tachyon inflation and canonical inflation differs by the value $a^2H^2\varepsilon_{1}\varepsilon_{2}$ [\citen{Steer:2003yu}].

\subsection{Inflation with Tachyon Scalar Field in RSII Model}

In tachyon inflation in RSII cosmology we have [\citen{Bilic:2016fgp,Bilic:2017yll,Dimitrijevic:2018nlc}]  
\begin{equation}
z = \frac{1}{\sqrt{4\pi G}}\frac{a\sqrt{\varepsilon_1}}{c_{\rm s}},
\end{equation}
where
\begin{equation}
c_{\rm s}^2=1-\frac{1}{3}\varepsilon_{1}.
\end{equation}
The parameters $n_{\rm s}$ and $r$ obtained for $A=1$ and $B=1/3$ are
\begin{equation}
n_{\rm s}-1=-2\varepsilon_{1}-\varepsilon_{2}-\left[2\varepsilon_{1}^{2}+\left(2C+\frac{17}{6}\right)\varepsilon_{1}\varepsilon_{2}+C\varepsilon_{2}\varepsilon_{3}\right],
\end{equation}
\begin{equation}
r=16\varepsilon_{1}[1+C\varepsilon_{2}-\frac{1}{6}\varepsilon_{1}].
\end{equation}
It may be easily verified that the same equation is obtained also from (\ref{nsst}) and (\ref{rst}) for $\alpha=1/12$. As in the previous case, the differences in parameters $n_{\rm s}$ and $r$ are affected by the different value of the sound of speed.

\subsection{Inflation with Tachyon Scalar Field in Holographic Model}

Perturbations in the tachyon model in holographic cosmology were first considered in [\citen{Bilic:2018uqx}] using an approximate scheme and the authors concluded that $z$ and $c_{\rm s}^2$ are given by (\ref{zhol}) and (\ref{cshol}), respectively. This gives rise to
\begin{eqnarray}
n_{\rm s} - 1 =
&-& \left(2+\frac{2h^2}{2-h^2}\right)\varepsilon_1 - \varepsilon_2
\nonumber\\
&-& \left(2+\frac{2h^2}{2-h^2} - \frac{8h^2}{3(4-h^2)^2} - \frac{8Ch^2}{(2-h^2)^2} \right)\varepsilon_1^2
\nonumber\\
&-& \left(\frac{8}{3} + \frac{h^2}{3(4-h^2)} + \frac{4C}{2-h^2}\right)\varepsilon_1\varepsilon_2 - C\varepsilon_2\varepsilon_3,
\end{eqnarray}
\begin{equation}
r= 8\left(2-h^2\right)\varepsilon_{1}\left[1+C\varepsilon_{2}+2\left(\frac{Ch^2}{2-h^2}-\frac{2-h^2}{12-3h^2}\right)\varepsilon_{1}\right].
\end{equation}
The same result is obtained using (\ref{n_s3}) and (\ref{finalr}) for $A=1-\frac{1}{2}h^2$ and $B=\frac{4(2-h^2)}{3(4-h^2)}$. The complete perturbation theory at linear order for the tachyon model in holographic cosmology, developed in [\citen{Bertini:2020dvv}], suggested that $z$ is of the form
\begin{equation}
z = \frac{1}{\sqrt{4\pi G}}\frac{a\sqrt{\varepsilon_1}}{c_{\rm s}},
\end{equation}
where $c_{\rm s}^2$ is given by (\ref{cshol}). Accepting $A=1$ and $B=\frac{4(2-h^2)}{3(4-h^2)}$  we reproduce the results from [\citen{Bertini:2020dvv}]
\begin{equation}
n_{\rm s}-1=-2\varepsilon_{1}-\varepsilon_{2}-\left[\left(2-\frac{8h^2}{3(4-h^2)^2}\right)\varepsilon_{1}^2-3\left(3+2C-\frac{2(2-h^2)}{3(4-h^2)}\right)\varepsilon_{1}\varepsilon_{2}-C\varepsilon_{2}\varepsilon_{3}\right],
\end{equation}
\begin{equation}
r=16\varepsilon_{1}\left[1+C\varepsilon_{2}-\frac{2(2-h^2)}{3(4-h^2)}\varepsilon_{1}\right].
\end{equation}

\subsection{Inflation with Canonical Scalar Field and Modified Friedmann Equation}

As a novelty, let extend the consideration to the inflationary model with canonical scalar field $\phi$ and the generalized (first)
Friedmann equation of the form 
\begin{equation}
{\cal F}(H)=\frac{8\pi G_{\rm N}}{3}\rho,
\end{equation}
and calculate the scalar spectral index and the tensor-to-scalar ratio up to the second order in the Hubble flow parameters. The parameters $n_{\rm s}$ and $r$ for this model are given at the lowest order in the slow-roll parameters $\epsilon_{H}$ and $\eta_{H}$ in Ref. [\citen{delCampo:2012qb}]
%

\begin{equation}
n_{\rm s}-1=2\eta_{H}-2\left(3-H\frac{{\cal F}_{,HH}}{{\cal F}_{,H}}\right)\epsilon_{H},
\label{nsdelC}
\end{equation}

\begin{equation}
r=2\frac{{\cal F}_{,H}}{H}\epsilon_{H},
\label{rdelC}
\end{equation}
where the subscript $,H$ denotes a partial derivative with respect to the Hubble parameter. The slow-roll parameters $\epsilon_{H}$ and $\eta_{H}$ are defined as follow [\citen{Liddle:1994dx}]
\begin{equation}
\epsilon_{H}\equiv -\frac{d\ln H}{d\ln a},
\label{defepsilon}
\end{equation}
\begin{equation}
\eta_{H}\equiv -\frac{d\ln H_{,\theta}}{d\ln a}.
\label{defeta}
\end{equation}
It is obvious that $\epsilon_{H}\equiv \varepsilon_{1}$. In order to conect  $\eta_{H}$ via $\varepsilon_{1}$ and $\varepsilon_{2}$ we use Hamilton-Jacobi formalism, i.e. $\dot{H}=H_{,\theta}\dot{\phi}$, and express  $\dot{\phi}$ with  [\citen{delCampo:2012qb}]
\begin{equation}
\dot{\phi}=-\frac{1}{8\pi G_{\rm N}}{\cal F}_{,H}\frac{H_{,\theta}}{H}.
\label{dotphi}
\end{equation}
From the definition for $\varepsilon_{2}$, using (\ref{dotphi}), we obtain
\begin{equation}
\eta_{H}=\frac{1}{2}\left(3-H\frac{{\cal F}_{,HH}}{{\cal F}_{,H}}\right)\varepsilon_{1}-\frac{1}{2}\varepsilon_{2}.
\end{equation}
Note that for  ${\cal F}(H)=H^2$ we recover the usual result $\eta=\varepsilon_{1}-1/2\varepsilon_{2}$ [\citen{Schwarz:2001vv}]. Now, (\ref{nsdelC}) and (\ref{rdelC}) can be recast in the form
\begin{equation}
n_{\rm s}-1=-\left[3-H\frac{{\cal F}_{,HH}}{{\cal F}_{,H}}\right]\varepsilon_{1}-\varepsilon_{2},
\label{nsdelC1}
\end{equation}

\begin{equation}
r=2\frac{{\cal F}_{,H}}{H}\varepsilon_{1}.
\label{rdelC1}
\end{equation}

In order to apply our approach  for calculating $n_{\rm s}$ and $r$ we need to find the function $A(H)$. Substituting (\ref{dotphi}) in $z$, defined in Ref.  [\citen{delCampo:2012qb}] as
\begin{equation}
z=a\frac{\dot{\phi}}{H},
\end{equation}
we obtain
\begin{equation}
z=\frac{1}{\sqrt{4\pi G_{\rm N}}}a\sqrt{\varepsilon_{1}}\sqrt{\frac{{\cal F}_{,H}}{2H}},
\end{equation}
and we observe that
\begin{equation}
A=\frac{{\cal F}_{,H}}{2H},\quad B=0.
\end{equation}
From  (\ref{n_s3}) and (\ref{finalr}) we find 
\begin{eqnarray}
n_{\rm s}-1=&-&\left[3-H\frac{{\cal F}_{,HH}}{{\cal F}_{,H}}\right]\varepsilon_{1}-\varepsilon_{2}\nonumber\\
&-&\left[3-C\left(H\frac{F_{,HH}}{F_{,H}}\right)^2+H\left(C\left(\frac{F_{,HH}}{F_{,H}}-H\frac{F_{,HHH}}{F_{,H}}\right)-\frac{F_{,HH}}{F_{,H}}\right)\right]\varepsilon_{1}^2\nonumber\\
&-&\left[3+C\left(3-H\frac{F_{,HH}}{F_{,H}}\right)\right]\varepsilon_{1}\varepsilon_{2}-C\varepsilon_{2}\varepsilon_{3},
\label{ourns}
\end{eqnarray}

\begin{equation}
r=8\frac{{\cal F}_{,H}}{H}\varepsilon_{1}\left[1+C\varepsilon_{2}+C \left(1-H\frac{F_{,HH}}{F_{,H}}\right)\varepsilon_{1}\right].
\label{ourr}
\end{equation}
Note that our results for $n_{\rm s}$ at linear order in $\varepsilon_{1}$ and $\varepsilon_{2}$ agree with (\ref{nsdelC1}), so we recover the result from Ref. [\citen{delCampo:2012qb}]. The difference in constant prefactor between expressions (\ref{rdelC1}) and (\ref{ourr}) is because in reference [\citen{delCampo:2012qb}] the tensor perturbation spectrum is not given in the same way as (\ref{PT}).  For ${\cal F}(H)=H^2$ expressions (\ref{ourns}) and (\ref{ourr}) coincide with (\ref{nsssfi}) and (\ref{rssfi}), as expected.

\section{Conclusions}

In this paper, by applying slow-roll approximation, we have obtained general expressions for the observational parameters of inflation, the scalar spectral index and the tensor-to-scalar ratio, for
several minimally coupled single scalar field models.  The results of the analysis of the first order cosmological perturbations, for single scalar field inflation minimally coupled with gravity in several cosmological scenarios,  motivated us to introduce a generalized form of the pump field $z$ and the speed of sound squared $c_{\rm s}^2$. Using those functions  we have calculated the generalized expression for the scalar power spectrum. By assuming the same form of the tensor power spectrum as in the single standard scalar field inflation the generalized expressions for  $n_{\rm s}$ and $r$, up to the second order in the Hubble flow parameters have been derived. Our calculation agree with the fact that the first order terms in the Hubble flow parameter expansion of $n_{\rm s}$ and $r$ depend on the pump field, while the second order terms depend additionally on the speed of sound. We have shown that the obtained expressions for the parameters give the same results as in canonical scalar field inflation (in the standard cosmology) and tachyon inflation (in the standard, RSII and holographic cosmology).  Additionally, adjusting our calculation to account the models of inflation with modified Friedmann equation we find $n_{\rm s}$ and $r$ at  quadratic order in the Hubble flow  parameters. 

It is natural to expect that  forthcoming observations will yield more rigorous constraints on the parameters $n_{\rm s}$ and $r$, demanding more precise analytical expressions that usually one, given at the lowest order in the slow-roll parameters. On the other hand, when the slow-roll parameters are not to small all quadratic terms in the slow-roll  parameters must be taken into account. We emphasize that calculated expressions for the parameters $n_{\rm s}$ and $r$ provide fast and elegant way to calculate the observational parameters at high accuracy in any other suitable models.

\section*{Acknowledgments}

This work has been supported by the ICTP-SEENET-MTP project NT-03 Cosmology-Classical and Quantum Challenges and the COST Action CA23130 "Bridging high and low energies in search of quantum gravity (BridgeQG)". M. Stojanovic acknowledges the support provided by The Ministry of Science, Technological Development and Innovation of the Republic of Serbia under contract 451-03-65/2024-03/200113. D. D. Dimitrijevic and M. Milosevic acknowledge the support provided by the same Ministry under contract 451-03-65/2024-03/200124. In addition, D. D. Dimitrijevic acknowledges the support of the CEEPUS Program RS-1514-03-2223 "Gravitation and Cosmology".



\section*{Appendix A}\label{appA}

In this Appendix we recall some important relation in slow-roll regime. To ensure a slow-roll regime the slow-roll parameters are introduced. In the present work, we use mainly the Hubble flow (slow-roll) parameters defined as
\begin{equation}
\varepsilon_{0}\equiv \frac{H_{*}}{H},
\label{cstachyon}
\end{equation}
\begin{equation}
\varepsilon_{i+1}\equiv \frac{d\ln|\varepsilon_{i}|}{dN},\quad i\geq 0,
\end{equation}
where $H_{*}$ is the Hubble parameter at some chosen time. The Hubble flow parameter, defined by (\ref{cstachyon}), have the property
\begin{equation}
\dot{\varepsilon}_{i}=\varepsilon_{i}\varepsilon_{i+1}H.
\end{equation}
The first three Hubble flow parameters are defined as
\begin{equation}
\varepsilon_1\equiv-\frac{\dot{H}}{H^2},
\label{e1}
\end{equation}
\begin{equation}
\varepsilon_2 \equiv \frac{\dot{\varepsilon}_1}{H\varepsilon_1},
\label{e2}
\end{equation}
\begin{equation}
\varepsilon_3 \equiv \frac{\dot{\varepsilon}_2}{H\varepsilon_2}.
\label{e3}
\end{equation}
During the slow-roll regime the Hubble flow parameters satisfy $\varepsilon_{i}\ll 1$,
and inflation ends when $\varepsilon_{1}$ exceeds unity.

In order to perform calculation we need expression for the conformal time $\eta$
\begin{equation}
\eta = \int\frac{1}{a}dt,
\label{eta_tau}
\end{equation}
in the slow-roll regime
\begin{equation}
\eta = -\frac{1}{aH}\frac{1-\varepsilon_1}{(1-\varepsilon_1)^2-\varepsilon_1\varepsilon_2} + \mathcal{O}(\varepsilon_{i}^3),
\label{eta_SR}
\end{equation}
which follows from the definition of $\varepsilon_{1}$ expressed in terms of the conformal time. We will use expression up to the first order in the Hubble flow parameters
\begin{equation}
\eta = -\frac{1}{aH}\frac{1}{1-\varepsilon_1} + \mathcal{O}(\varepsilon_{i}^2)
\simeq -\frac{1}{aH}(1+\varepsilon_1) + \mathcal{O}(\varepsilon_{i}^2),
\label{eta_SR1}
\end{equation}
and its derivative with respect to time $t$
\begin{equation}
\frac{d\eta }{dt }= \frac{1}{a}(1-\varepsilon_1^2-\varepsilon_1\varepsilon_2) + \mathcal{O}(\varepsilon_{i}^3).
\label{deta_dtau}
\end{equation}
\noindent Using (\ref{e1})-(\ref{e3}) and (\ref{deta_dtau}) we can write useful expressions for further calculations
\begin{eqnarray}
\dot{H} &=& -H^2\varepsilon_1,
\label{h_dot}\\
\dot{\varepsilon}_1 &=& H\varepsilon_1\varepsilon_2,
\label{e1_dot}\\
\dot{\varepsilon}_2 &=& H\varepsilon_2\varepsilon_3,
\label{e2_dot}\\
\frac{dt }{d\eta } &=&
\frac{a}{1-\varepsilon_1^2-\varepsilon_1\varepsilon_2} + \mathcal{O}(\varepsilon_{i}^3),
\label{dtau_deta}\\
\left(\frac{dt }{d\eta }\right)^2 &=&
\frac{a^2}{(1-\varepsilon_1^2-\varepsilon_1\varepsilon_2)^2} + \mathcal{O}(\varepsilon_{i}^3),
\label{dtau_deta1}\\
\frac{d}{dt}\left(\frac{dt }{d\eta }\right) &=&
\frac{aH}{1-\varepsilon_1^2-\varepsilon_1\varepsilon_2} + \mathcal{O}(\varepsilon_{i}^3),
\label{dtau_deta2}\\
\frac{d}{dt}\left(\frac{dt }{d\eta }\right)\frac{dt }{d\eta } &=&
\frac{a^2H}{(1-\varepsilon_1^2-\varepsilon_1\varepsilon_2)^2} + \mathcal{O}(\varepsilon_{i}^3).
\label{dtau_deta3}
\end{eqnarray}
\noindent Similarity of expressions (\ref{dtau_deta1}) and (\ref{dtau_deta3}) leads to
\begin{equation}
\frac{d}{dt}\left(\frac{dt }{d\eta }\right)\frac{dt }{d\eta } =
H\left(\frac{dt }{d\eta }\right)^2.
\label{dtau_deta4}
\end{equation}

\section*{Appendix B}\label{appB}

In this Appendix we explicitly calculate the coefficients $b_i$ in (\ref{9}). To do that and to obtain the whole expression for $\nu$ we need to rewrite (\ref{z_second_over_z_prefinal1}) with the help of ($\ref{eta_SR1}$) as
\begin{eqnarray}
\frac{z^{\prime\prime}}{z} &=& \frac{1}{\eta^2}(1+\varepsilon_1)^2
\left\{
2 - \left[1+\frac{3}{2}H\left(\frac{1}{A}\frac{dA}{dH}\right)\right]\varepsilon_1 + \frac{3}{2}\varepsilon_2\right\}
\nonumber\\
&+&
\frac{1}{\eta^2}(1+\varepsilon_1)^2
\left\{
\left[4+A_1\right]\varepsilon_1^2 + \left[2+A_2\right]\varepsilon_1\varepsilon_2 + \frac{1}{4}\varepsilon_2^2 +
\frac{1}{2}\varepsilon_2\varepsilon_3 + \mathcal{O}(\varepsilon_{i}^3)
\right\}.
\label{z_second_over_z_final}
\end{eqnarray}
\noindent After some rearrangements we can write
\begin{eqnarray}
\frac{z^{\prime\prime}}{z} &=& \frac{1}{\eta^2}
\left\{
2 + \left[3-\frac{3}{2}H\left(\frac{1}{A}\frac{dA}{dh}\right)\right]\varepsilon_1 + \frac{3}{2}\varepsilon_2\right\}
\nonumber\\
&+&
\frac{1}{\eta^2}
\left\{
\left[4+A_1-\frac{3}{2}H\left(\frac{1}{A}\frac{dA}{dH}\right)\right]\varepsilon_1^2 +
\left[5+A_2\right]\varepsilon_1\varepsilon_2 + \frac{1}{4}\varepsilon_2^2 +
\frac{1}{2}\varepsilon_2\varepsilon_3 + \mathcal{O}(\varepsilon_{i}^3)
\right\}.
\label{z_second_over_z_final1}
\end{eqnarray}
Comparing (\ref{z_second_over_z_prefinal2}) and (\ref{z_second_over_z_final1}) we see that
\begin{eqnarray}
\nu^2 &=& \frac{9}{4} + \left[3-\frac{3}{2}H\left(\frac{1}{A}\frac{dA}{dH}\right)\right]\varepsilon_1 + \frac{3}{2}\varepsilon_2
\nonumber\\
&+&
\left[4+A_1-\frac{3}{2}H\left(\frac{1}{A}\frac{dA}{dH}\right)\right]\varepsilon_1^2 +
\left[5+A_2\right]\varepsilon_1\varepsilon_2 + \frac{1}{4}\varepsilon_2^2 +
\frac{1}{2}\varepsilon_2\varepsilon_3 + \mathcal{O}(\varepsilon_{i}^3).
\label{nu_squared}
\end{eqnarray}

We can now put $\nu$ in general form (as we formally did writing expression (\ref{9}))
\begin{eqnarray}
\nu = b_0 + b_1\varepsilon_1 + b_2\varepsilon_2 +
b_3\varepsilon_1^2 + b_4\varepsilon_1\varepsilon_2 + b_5\varepsilon_2^2 +
b_6\varepsilon_2\varepsilon_3.
\label{nu}
\end{eqnarray}
Squaring (\ref{nu}) and keeping terms up to the second order gives
\begin{eqnarray}
\nu^2 &=& b_0^2 + 2b_0b_1\varepsilon_1 + 2b_0b_2^2\varepsilon_2 +( b_1^2+2b_0b_3)\varepsilon_1^2 
\nonumber\\
&+&
(2b_0b_4+2b_1b_2)\varepsilon_1\varepsilon_2 + (b_2^2+2b_0b_5)\varepsilon_2^2 +
2b_0b_6\varepsilon_2\varepsilon_3 + \mathcal{O}(\varepsilon_{i}^3).
\label{nu_squared1}
\end{eqnarray}
Comparing (\ref{nu_squared}) and (\ref{nu_squared1}) we deduce
\begin{eqnarray}
&& b_0=\frac{3}{2}, \quad b_1=1-\frac{1}{2}H\left(\frac{1}{A}\frac{dA}{dH}\right),\quad b_2=\frac{1}{2},
\nonumber\\
&&
b_3=\frac{1}{3}\left\{\left[4+A_1-\frac{3}{2}H\left(\frac{1}{A}\frac{dA}{dH}\right)\right]-
\left[1-\frac{1}{2}H\left(\frac{1}{A}\frac{dA}{dH}\right)\right]^2\right\},
\nonumber\\
&&
b_4=\frac{1}{3}\left[4+A_2+\frac{1}{2}H\left(\frac{1}{A}\frac{dA}{dH}\right)\right], \quad b_5=0,
\quad b_6=\frac{1}{6}.
\label{nu_squared2}
\end{eqnarray}
Finally, we can write
\begin{eqnarray}
\nu &=& \frac{3}{2} + \left[1-\frac{1}{2}H\left(\frac{1}{A}\frac{dA}{dH}\right)\right]\varepsilon_1 +
\frac{1}{2}\varepsilon_2
\nonumber\\
&+&
\frac{1}{3}\left\{\left[4+A_1-\frac{3}{2}H\left(\frac{1}{A}\frac{dA}{dH}\right)\right]-
\left[1-\frac{1}{2}H\left(\frac{1}{A}\frac{dA}{dH}\right)\right]^2\right\}\varepsilon_1^2
\nonumber\\
&+&
\frac{1}{3}\left[4+A_2+\frac{1}{2}H\left(\frac{1}{A}\frac{dA}{dH}\right)\right]\varepsilon_1\varepsilon_2 +
\frac{1}{6}\varepsilon_2\varepsilon_3,
\label{nu_final}
\end{eqnarray}
where, the expressions for $A_1$ and $A_2$ are given in (\ref{B1}) and (\ref{B2}).


\end{document}